# Cerebral microbleeds: Association with cognitive decline and pathology build-up

Saima Rathore[1,2], Jatin Chaudhary[3], Boning Tong[4], Selen Bozkurt[1]

[1]Department of Biomedical Informatics, Emory University School of Medicine, Atlanta, GA, USA, 30329

[2]Department of Neurology, Emory University School of Medicine, Atlanta, GA, USA, 30329

[3]University of Turku, Turku, Finland, FI-20014

[4]Department of Bioengineering, School of Engineering and Applied Science, University of Pennsylvania, Philadelphia, PA, USA, 19104

## Abstract

**Introduction:** Cerebral microbleeds are hypothesized downstream markers of brain damage caused by vascular and amyloid pathologic mechanisms. Cerebral microbleeds are associated with deteriorated cognition in aging individuals, however, it is unknown how microbleeds are related to the onset and progression in Alzheimer Disease (AD). The purpose of this analysis was to test whether the presence and location of lobar microbleeds demonstrate differential associations with amyloid-β (Aβ)-PET, tau tangle formation (tau-PET), and longitudinal cognitive decline.

**Design, Setting, and Participants**: A total of 1,573 subjects from the Alzheimer's Disease Neuroimaging Initiative (ADNI), who had undergone 3T MR imaging and had information on the number and location of microbleeds assessed visually, were included in the analysis. A subset of these patients also underwent lumbar puncture, and Aβ-PET and tau-PET scans were acquired. Associations between lobar microbleeds and pathology (Aβ-PET and tau-PET levels), cerebrospinal fluid (CSF), genetics, and cognition were evaluated at the granular level, focusing on regional microbleeds and pathology, as well as domain-specific cognitive decline, using ordinary least-squares regression while adjusting for covariates. The analysis was repeated to assess associations with longitudinal cognitive decline. Cognitive decline was measured using the Alzheimer's Disease Assessment Scale-Cognition (ADAS-Cog11), and corresponding domain-specific sub-scores were calculated. Participants underwent neuropsychological testing at least twice, with a minimum interval of two years between assessments.

**Results:** A total of 1,573 participants (692 women [43.9%]; mean [SD] age, 71.23 [7.19] years) underwent baseline and follow-up cognitive testing. The prevalence of microbleeds was 24.0% (373 participants). The presence of microbleeds was associated with cognitive decline; microbleeds in the temporal lobe were linked to declines in the semantic (coefficient = 0.73; 95% CI, 0.02 to 0.10; P=.004), language (coefficient = 0.16; 95% CI, 0.03 to 0.28; P=0.013), and praxis domains (coefficient = 0.09; 95% CI, 0.00 to 0.18; P=0.039). Additionally, microbleeds in the overall cortex were associated with a decline in the language domain (coefficient = 0.10; 95% CI, 0.02 to 0.18; P = 0.018). While assessing pathology, we found that the presence of microbleeds in the temporal lobe was associated with tau increase in overall cortex. Similarly, microbleeds in the overall cortex were associated with increased Aβ-PET in temporal, parietal and frontal regions of the brain.

**Conclusions**: In a mixed population, presence of overall microbleeds was associated with longitudinal cognitive decline specifically in semantic and language domains and was associated with higher baseline Aβ and tau pathology. Results suggest incorporating lobar information of microbleeds into the diagnostic and prognostic workup of AD.

## 1. Introduction

Cerebral microbleeds (CMB) are small, circular focal hemorrhages that can be detected through T2*-weighted gradient-recalled echo (GRE) or susceptibility-weighted imaging (SWI) on magnetic resonance imaging (MRI). These microbleeds are often linked to small vessel arteriosclerosis, particularly in individuals with hypertension, as well as cerebral amyloid angiopathy (CAA). Their location can provide insight into the underlying causes [1]. In healthy older adults, the estimated prevalence of CMB ranges from 5% to 15%, and they have been associated with a heightened risk of cognitive decline and dementia [2]. Hypertensive individuals were found to be four times more likely to have microbleeds compared to the general population [3], especially when accompanied by other indicators of small-vessel disease, such as white matter hyperintensities and lacunar infarcts [4], [5]. A meta-analysis conducted in 2016, which reviewed 15 MRI studies, found that CMB were observed in roughly 24% of

Alzheimer's disease (AD) patients; however, the reported prevalence met with some skepticism (Sepehry et al., 2016).

MRI- and positron emission tomography (PET)-based markers including the traditional standardized uptake volume ratio (SUVr) values from the atlas-based regions of PET images or the advanced texture features and atrophy patterns from MRI have been associated with clinical diagnosis and disease progression [6], [7], [8], [9]. Similarly, genomic markers have also been associated with pathology and cognitive impairment over time [10], [11], [12], [13], [14], however, there is little and mixed literature on the role of microbleeds in AD onset and progression. Microbleeds can occur throughout the brain, and their location may provide clues about the underlying pathology. In patients with hypertension, microbleeds are commonly found in the deep gray matter, often resulting from hemosiderin leakage due to abnormalities in small blood vessels [15]. In contrast, microbleeds in older individuals and those with Alzheimer's disease (AD) are predominantly lobar, typically associated with cerebral amyloid angiopathy, where amyloid-β deposits accumulate along the walls of blood vessels, leading to vessel fragility and leakage [16]. Understanding the association between cerebral microbleeds, tau, amyloid, and cognitive decline in Alzheimer's disease (AD) is crucial for unraveling the interplay between vascular pathology and hallmark AD biomarkers. Microbleeds, often linked to cerebral amyloid angiopathy (CAA), may exacerbate the deposition of amyloid and tau proteins, contributing to faster cognitive decline. Investigating these associations will provide insight into how vascular abnormalities influence the progression of AD, aiding in the identification of high-risk individuals and informing future therapeutic strategies.

The purpose of our analysis was to determine whether the distribution of lobar microbleeds has differential associations with downstream events in AD pathogenesis such as pathology buildup and cognitive decline. Specifically, by using data from the multicenter Alzheimer's Disease Neuroimaging Initiative (ADNI) [17], we tested the hypotheses that lobar microbleeds 1) are associated with brain amyloidosis, reflected by higher amyloid-PET levels, 2) are related to tau tangle burden, reflected by higher tau-PET levels, and 3) predict greater longitudinal cognitive decline. Since the microbleeds in the setting of AD are mainly lobar [3], [18] we used only lobar microbleeds in this analysis.

## 2. Materials and Methods

### 2.1 Subjects

This study analyzed a cohort of 1,573 participants from the ADNI dataset, all of whom had undergone 3T MRI and cognitive assessments. A subset of these participants also underwent PET scans to assess Aβ plaques and tau neurofibrillary tangles using Aβ-PET and tau-PET imaging. ADNI is an ongoing longitudinal, multicenter observational study designed to discover imaging and biochemical biomarkers for diagnosing and tracking Alzheimer's disease [19]. The study received approval from the institutional review boards of all participating sites, with written informed consent obtained from each participant.

Subjects enrolled in later phases of ADNI, including ADNI-2, ADNI–Grand Opportunity, and ADNI-3, underwent T2* gradient recalled-echo sequences that was used to detect the number and location of microbleeds in the brain. These data were accessed through the LONI database, which houses ADNI's publicly available dataset. Participants were aged between 55 and 90 years, with eligibility criteria requiring no significant neurological or psychiatric conditions, systemic illnesses, or lab abnormalities that could affect follow-up. Additionally, to reduce the potential influence of comorbid vascular disease, subjects were included only if they had a low modified Hachinski ischemic score of <=4 [20]

### 2.2 Cognitive assessment and domains

At baseline, all participants completed the Alzheimer's Disease Assessment Scale-Cognitive (ADAS-Cog) evaluation [21], the most widely used measure in clinical trials. The ADAS-Cog assesses episodic and semantic memory, language, executive function, and praxis, with higher scores indicating worse performance in each domain. This cognitive assessment was repeated longitudinally. Sub-scores for episodic memory were calculated using items from the ADAS-Cog related to recall, recognition, orientation, and remembering instructions. Semantic memory was measured using the naming objects and fingers sub-scores, while the language sub-score included difficulty in finding words and following commands. Finally, the ideational and constructional praxis sub-scores were used to assess the praxis domain.

## 2.3 MR Imaging Processing

Each subject underwent standardized 3T MR imaging protocol, which included T1-weighted 3D MPRAGE images and T2* gradient recalled-echo images. More details on imaging protocol can be found on ADNI LONI web portal. http://adni.loni.usc.edu/methods/documents/mri-protocols/. This standardized imaging protocol included the following parameters for the gradient recalled-echo sequence: TE, 20 ms; TR, 650 ms; flip angle, 20°; section thickness, 4 mm; section gap, 0 mm.

T2*-weighted images are sensitive to micro-hemorrhages (MCH) and superficial siderosis, which appear as small dark spots in the images. These images, acquired in 3 mm axial slices with 1 mm in-plane resolution, were used to identify microbleeds—hypointense lesions smaller than 10 mm. Only definite microbleeds were analyzed in this study. Microbleeds were classified by location as either deep gray matter/infratentorial (basal ganglia, thalami, brainstem, or cerebellum) or lobar (other brain regions of the brain parenchyma), however, this study focused solely on lobar microbleeds.

## 2.4 PET imaging Processing

Preprocessed images from ADNI, with realigned frames, head position corrected through linear transformation, standardized voxel size, and smoothed to a uniform resolution of 6mm, were used in this study. Florbetapir PET scans were analyzed in each participant's native space, using their structural MRIs acquired closest to the florbetapir PET scan. The structural MRIs were segmented into cortical regions of interest and reference regions for each subject using FreeSurfer. The florbetapir PET data were then realigned, and the mean of all frames was used to co-register the florbetapir data with the corresponding structural MRI.

For each subject and time point, cortical standardized uptake value ratio (SUVR) images were generated by dividing voxel-wise florbetapir uptake by the average uptake from the whole cerebellum reference region. Florbetapir SUVR values were quantified within each FreeSurfer Desikan-Killiany region (left, right, and volume-weighted bilateral). Using these regional summary statistics, average cortical florbetapir SUVR values were calculated for each brain lobe by combining the respective values from the relevant brain regions.

For tau-PET data, a similar procedure was followed. The tau-PET images were realigned, and the mean of all frames was used to co-register the tau-PET data with the structural MRI closest in time. In each subject's native MRI space, tau-PET SUVR images were generated by normalizing mean tau-PET uptake to a grey matter-masked cerebellum reference region.

## 2.5 CSF Biomarkers

A subset of patients in this analysis underwent lumbar puncture to collect cerebrospinal fluid (CSF) samples, allowing for the measurement of Aβ1-42, tau, and phosphorylated tau 181 (p-tau) [22]. Extensive research shows that the proteomic composition of CSF and plasma often mirrors brain-related pathophysiology. AD diagnostics now routinely include CSF biomarkers such as Aβ1-42, tau, and p-tau, which reflect the proteinopathies characteristic of the disease [23]. This study focuses on p-tau 181, total tau, and Aβ1-42 in the analysis.

## 2.6 Apolipoprotein E Genotyping

Each participant underwent apolipoprotein E (ApoE) genotyping at the baseline visit. Approximately 6 mL of blood was obtained from each participant in an ethylenediamine tetraacetic acid tube, gently mixed by inversion, and shipped at ambient temperature to a single designated laboratory within 24 hours of collection for genotyping analysis. ApoE ε4 status is defined as non-carriers, having no copy of ε4 allele, heterozygous, having one copy of ε4 allele, and homozygous, having two copied of ε4 allele.

## 2.7 Statistical Analysis

All statistical analyses were performed using R version 4.4.1 (2024-06-14) (StataCorp, College Station, Texas). In the cross-sectional analysis, we aimed to test the hypothesis that lobar microbleeds are associated with global cognitive function as well as specific cognitive subdomains measured by the ADAS-Cog. To assess this, we employed ordinary least-squares (OLS) regression models. The predictor variable was the presence of lobar microbleeds, dichotomized into two categories: participants with no microbleeds and those with one or more than one microbleed. The outcome variables included global cognitive scores and specific domain scores from the ADAS-Cog.

We adjusted for key covariates: age, sex, years of education, ApoE ε4 genotype, and diagnostic group (healthy controls, MCI, or AD).

To further explore the effect of microbleeds, we tested the relationship between lobar microbleeds and AD pathology, we repeated the regression analysis to assess the association between lobar microbleeds and levels of tau and amyloid pathology, as measured by Aβ- and tau-PET scans. Similar OLS models were used, controlling for the same covariates (age, sex, education, ApoE ε4 genotype, and diagnostic group). In these analyses, the presence of lobar microbleeds (dichotomized as above) was used as the predictor variable, while global amyloid and tau levels served as the outcome variables. Additionally, we conducted region-specific analyses to investigate whether microbleeds in particular lobes were associated with tau or amyloid pathology in the corresponding or other lobes of the brain. Finally, we tested whether microbleeds have association with CSF proteomic markers including total tau, p-Tau and Aβ1-42.

For the longitudinal analysis, we sought to determine whether lobar microbleeds were associated with cognitive decline over time. We used OLS regression models with the same covariates (age, sex, years of education, ApoE ε4 genotype, and diagnostic group). The annualized-rate-of-change in cognition was quantified by subtracting baseline score from the follow-up and dividing it with the exact number of years passed between them. The predictor variable was the dichotomous microbleed category, while the outcome variables was the rate of global and domain-specific cognitive decline. This allowed us to assess whether the presence of lobar microbleeds contributed to a faster decline in cognition over time.

## 3. Results

Subject characteristics are presented in Table 1. Of the 1573 subjects, 1200 (76%) had no microbleeds, whereas 373 (24%) had at least one microbleed. Of the 373 subjects with microbleeds, 333 (89%) had at least one lobar microbleed, and 40 (11%) had only deep gray/infratentorial microbleeds. ApoE ε4 carriership status led to increase in the number of microbleeds with ApoE ε4 carries having two copies of alleles had the highest number of microbleeds.

**Table 1:** Baseline group characteristics

| Characteristics | Microbleeds Absent | Microbleeds present |
|---|---|---|
| No. of subjects | 1200 (76%) | 373 (24%) |
| Mean age (SD), y | 70.92 (7.17) | 74.50 (7.22) |
| Sex, female/male | 561/508 | 131/174 |
| Mean education (SD), y | 16.35 (2.48) | 16.12 (2.70) |
| Race:White/Others | 909/160 | 275/30 |
| APOE4 Carrier Status | | |
| Non-Carriers | 536 | 138 |
| Homozygous | 328 | 98 |
| Heterozygous | 71 | 38 |
| Clinical Diagnosis | | |
| # AD | 123 | 63 |
| # LMCI | 281 | 60 |
| # EMCI | 189 | 87 |
| # SMC | 253 | 38 |
| # NC | 216 | 54 |
| Fluid Biomarkers | | |
| Mean FDG (SD) | 1.22 (0.16) | 1.17 (0.15) |
| Mean ABETA | 913.68 | 788 |
| Mean TAU | 269.35 | 316.65 |
| Mean PTAU | 25.49 | 30.86 |

Note: LMCI, Late mild cognitive impairment; EMCI, Early mild cognitive impairment; SMC=Subjective memory complaints; AD=Alzheimer's Disease; SD, standard deviation.

### 3.1 Topographic distribution of cerebral microbleeds

The topographic distribution of cerebral microbleeds included 81 of 333 participants (24.3%) with at least 1 microbleed in the occipital lobe, 101 (30.0%) in the parietal lobe, 83 (24.9%) in the temporal lobe, 128 (38.4%) in the frontal lobe. More detailed characterization is given in Figure 1.

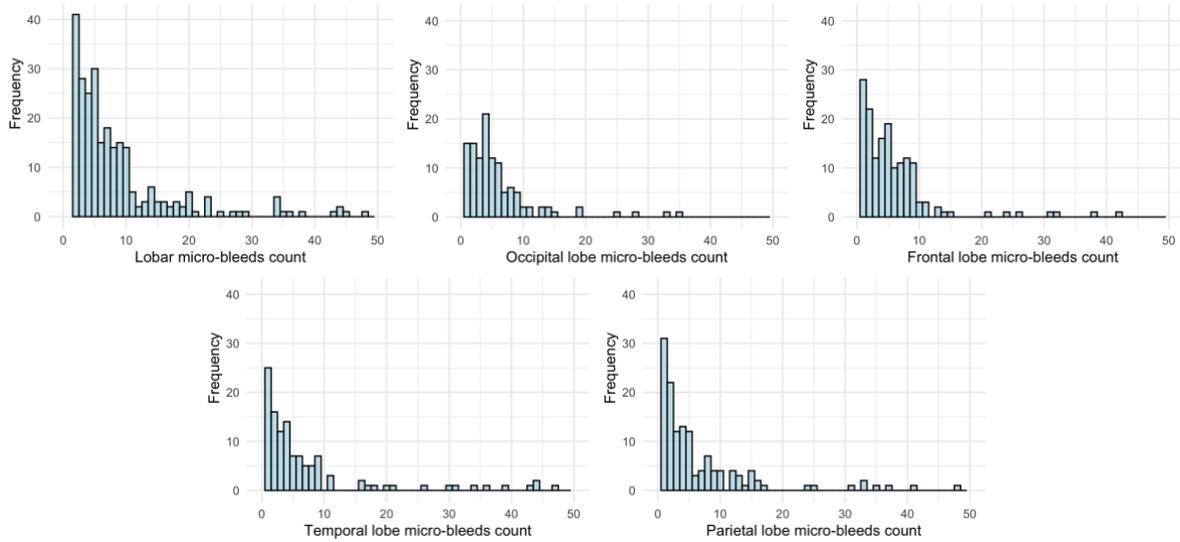

Figure 1: Topographic distribution of cerebral microbleeds in whole cortex and in different lobar regions (frontal, parietal, occipital, and temporal) of the brain.

### 3.2 Lobar Microbleeds and Genetic markers

The presence of the ApoE ε4 allele was correlated with an increased average number of cerebral microbleeds among the participants. The average number of microbleeds exhibited a positive correlation with the number of ApoE ε4 allele copies, rising from 0 to 2 copies, respectively shown as ε4-, ε4+ and ε4++. This trend was observed both at a global level and within specific brain lobes (average number of cerebral microbleeds, overall brain: ε4- = 3.89, ε4+ = 8.01, ε4++ = 27.18; occipital: ε4- = 0.85, ε4+ = 1.36, ε4++ = 4.10; temporal: ε4- = 0.84, ε4+ = 1.48, ε4++ = 7.94; frontal: ε4- = 1.21, ε4+ = 3.11, ε4++ = 8.86; parietal: ε4- = 0.97, ε4+ = 2.05, ε4++ = 6.27).

### 3.3 Lobar Microbleeds and Cognitive Performance

Compared with no microbleeds, the presence of any microbleed was not associated with a worse performance in cognition as measured by global ADAS-Cog11 score (coefficient=0.27, 95%CI: -0.30 to 0.83, P = 0.355) and domain-specific sub-scores (episodic memory: coefficient=-0.04, 95%CI: -0.21 to 0.13, P = 0.632; semantic memory: coefficient=0.03, 95%CI: -0.02 to 0.09, P = 0.242; language: coefficient=-0.01, 95%CI: -0.11 to 0.10, P = 0.909; praxis: coefficient=0.00, 95% CI: -0.10 to 0.10, P = 0.983). Similarly, when evaluated regionally, no significant association was observed between the cognitive performance in any domain and the presence of microbleeds in different lobes of the brain.

**Table 2:** Ordinary least square regression model demonstrating association between lobar microbleeds and likelihood of domain-specific cognitive impairment at baseline. Values given are effect size (95% confidence interval) p-value.

| Cognitive tests/domains | Microbleed present/absent? | | | | |
|---|---|---|---|---|---|
| | Overall brain | Parietal Lobe | Occipital Lobe | Frontal Lobe | Temporal Lobe |
| Global ADAS11 | 0.27 (-0.30 to 0.83) 0.355 | 4.97 (-0.40 to 1.23) 0.321 | 0.30 (-0.83 to 0.88) 0.954 | 4.30 (-0.36 to 1.08) 0.331 | 0.23 (-0.83 to 0.87) 0.965 |
| Episodic | -0.04 (0.21 to 0.13) 0.632 | -0.52 (-0.29 to 0.20) 0.726 | -1.18 (-1.21 to 1.01) 0.862 | -0.22 (-0.23 to 0.20) 0.867 | -0.99 (-0.34 to 0.17) 0.525 |
| Semantic | 0.03 (0.02 to 0.09) 0.242 | -0.22 (-0.10 to 0.06) 0.672 | 0.69 (-0.03 to 0.14) 0.194 | -0.15 (-0.09 to 0.06) 0.744 | -0.00 (-0.09 to 0.09) 0.999 |
| Language | -0.01 (-0.11 to 0.10) 0.909 | -0.03 (-0.19 to 0.12) 0.691 | -0.02 (-0.18 to 0.14) 0.831 | 0.08 (-0.06 to 0.21) 0.267 | 0.09 (-0.07 to 0.25) 0.251 |
| Praxis | -0.00 (-0.10 to 0.10) 0.983 | 0.11 (-0.04 to 0.260 0.144 | 0.03 (-0.13 to 0.18) 0.732 | -0.05 (-0.19 to 0.08) 0.415 | -0.05 (-0.20 to 0.11) 0.564 |

### 3.4 Lobar Microbleeds and Cognitive Decline

While evaluating longitudinally, the presence of overall microbleeds was not associated with global cognitive decline (coefficient=0.22, 95% CI: -0.09 to 0.96, P = 0.543) and decline in episodic (coefficient =0.05, 95% CI, -0.15 to 0.26, P = 0.621), semantic (coefficient = 0.02, 95% CI, -0.01 to 0.05, P = 0.125), and praxis domains (coefficient = 0.05, 95% CI, -0.01 to 0.10, P = 0.107). We did, however, observe that the presence of microbleeds was associated with faster cognitive decline in the language domain (coefficient = 0.10, 95% CI, 0.02 to 0.18, P = 0.018).

When evaluated regional microbleeds, no significant association was observed between the presence of microbleeds in any lobe with global cognitive decline and the decline in episodic and praxis domain except association found between temporal microbleeds and cognitive decline in praxis domain (coefficient = 0.09, 95% CI, 0.00 to 0.18, P = 0.039). We did, however, observe that the presence of microbleeds in various lobes was associated with faster decline in semantic (parietal: coefficient=0.53, 95% CI, 0.01 to 0.08, P = 0.027; occipital: coefficient = 0.50, 95% CI, 0.00 to 0.08, P = 0.048; temporal: coefficient = 0.73, 95% CI, 0.02 to 0.10, P = 0.004) and language domains (parietal: coefficient = 0.13, 95% CI, 0.01 to 0.25, P = 0.031; occipital: coefficient = 0.14, 95% CI= 0.02 to 0.26, P = 0.026; temporal: coefficient=0.16, 95% CI=0.03 to 0.28, P = 0.013).

**Table 3:** Ordinary least square regression model demonstrating association between lobar microbleeds and likelihood of domain-specific cognitive decline (longitudinal change in cognition). Values given are effect size (95% confidence interval) p-value. Bold entries show significant associations.

| Cognitive tests/domains | Microbleed present/absent? | | | | |
| --- | --- | --- | --- | --- | --- |
| | Overall brain | Parietal Lobe | Occipital Lobe | Frontal Lobe | Temporal Lobe |
| Global ADAS11 | 0.22 (-0.09 to 0.52) 0.160 | 3.48 (-0.15 to 0.73) 0.199 | 2.76 (-0.23 to 0.69) 0.330 | 0.88 (-0.32 to 0.47) 0.713 | 4.59 (-0.09 to 0.85) 0.110 |
| Episodic | 0.05 (-0.15 to 0.26) 0.621 | 1.01 (-0.22 to 0.38) 0.582 | 0.73 (-0.25 to 0.37) 0.705 | 0.08 (-0.26 to 0.27) 0.963 | -0.26 (-0.34 to 0.30) 0.893 |
| Semantic | 0.02 (-0.01 to 0.05) 0.125 | **0.53 (0.01 to 0.08) 0.027** | **0.50 (0.00 to 0.08) 0.048** | 0.15 (-0.02 to 0.05) 0.495 | **0.73 (0.02 to 0.10) 0.004** |
| Language | **0.10 (0.02 to 0.18) 0.018** | **0.13 (0.01 to 0.25) 0.031** | **0.14 (0.02 to 0.26) 0.026** | -0.00 (-0.11 to 0.10) 0.963 | **0.16 (0.03 to 0.28) 0.013** |
| Praxis | 0.05 (-0.01 to 0.10) 0.107 | 0.08 (-0.00 to 0.16) 0.062 | 0.05 (-0.04 to 0.14) 0.275 | 0.01 (-0.06 to 0.08) 0.805 | **0.09 (0.00 to 0.18) 0.039** |

### 3.5 Lobar Microbleeds Are Associated with Tau Accumulation

Compared with no microbleeds, the presence of lobar microbleeds was associated with higher tau-PET signal in the cortex (coefficient = 0.07, 95% CI=0.01 to 0.13, P = 0.029). Similarly, when evaluated in different lobes of the brain, presence or absence of microbleeds was associated with higher tau-PET SUVR in frontal lobe (coefficient = 0.06, 95% CI=0.01 to 0.12, P = 0.024), and temporal lobe (coefficient = 0.08, 95% CI= 0.00 to 0.16, P = 0.053).

When microbleeds were assessed in different lobes of the brain, the presence of absence of microbleeds in parietal, occipital, and frontal lobes was not associated with higher tau-PET signal in any lobe of the brain. However, the presence or absence of microbleeds in temporal lobe was associated with higher tau-PET SUVR in meta temporal ROI (coefficient = 0.21, 95% CI=0.05 to 0.37, P=0.009), parietal lobe (coefficient = 0.23, 95% CI=0.08 to 0.38, P=0.002), frontal lobe (coefficient=0.15, 95% CI=0.06 to 0.24, P=0.001), occipital lobe (coefficient = 0.13, 95% CI=-0.02 to 0.27, P=0.100), temporal lobe (coefficient = 0.19, 95% CI= 0.06 to 0.33, P=0.006) and whole cortex (coefficient=0.18, 95% CI= 0.07 to 0.29, P=0.001).

**Table 4:** Ordinary least square regression model demonstrating association between lobar microbleeds and tau-PET SUVr in lobar regions of the brain at baseline. Values given are effect size (95% confidence interval) p-value. Bold entries show significant associations.

| Tau-PET SUVr | Presence of microbleeds? | | | | |
| --- | --- | --- | --- | --- | --- |
| | All microbleeds | Parietal microbleeds | Occipital microbleeds | Frontal microbleeds | Temporal microbleeds |
| Cortex | **0.07 (0.01 to 0.13) 0.029** | 0.02 (-0.08 to 0.12) 0.730 | -0.01 (-0.12 to 0.11) 0.896 | 0.05 (-0.05 to 0.14) 0.329 | **0.18 (0.07 to 0.29) 0.001** |
| Temporal | **0.08 (-0.00 to 0.16) 0.053** | 0.07 (-0.05 to 0.20) 0.247 | -0.06 (-0.20 to 0.09) 0.430 | 0.05 (-0.07 to 0.17) 0.395 | **0.19 (0.06 to 0.33) 0.006** |
| Occipital | 0.06 (-0.03 to 0.14) 0.184 | -0.01 (-0.14 to 0.13) 0.893 | -0.05 (-0.20 to 0.11) 0.531 | 0.09 (-0.04 to 0.22) 0.170 | **0.13 (-0.02 to 0.27) 0.100** |
| Frontal | **0.06 (0.01 to 0.12) 0.024** | -0.00 (-0.09 to 0.08) 0.916 | 0.04 (-0.05 to 0.14) 0.400 | 0.03 (-0.05 to 0.11) 0.438 | **0.15 (0.06 to 0.24) 0.001** |
| Parietal | 0.08 (-0.00 to 0.17) 0.060 | 0.01 (-0.12 to 0.14) 0.872 | -0.04 (-0.19 to 0.12) 0.642 | 0.04 (-0.09 to 0.17) 0.518 | **0.23 (0.08 to 0.38) 0.002** |
| Meta temporal | 0.09 (-0.01 to 0.18) 0.068 | 0.09 (-0.06 to 0.23) 0.231 | -0.08 (-0.24 to 0.08) 0.346 | 0.06 (-0.07 to 0.20) 0.372 | **0.21 (0.05 to 0.37) 0.009** |

### 3.6 Lobar Microbleeds Are Associated with Amyloid Accumulation

Compared with no microbleeds, the presence of lobar microbleeds in the whole cortex was associated with higher amyloid deposition in the whole cortex (coefficient = 0.04, 95% CI= 0.00 to 0.08, P = 0.041), temporal lobe (coeffcieint = 0.04, 95% CI=0.00 to 0.07, P = 0.048), frontal lobe (coefficient = 0.04, 95% CI = 0.00 to 0.08, P = 0.047), and parietal lobe (coefficient = 0.04, 95% CI = 0.00 to 0.08, P = 0.043). Microbleeds in any lobe of the brain were not associated with amyloid deposition in any region of the brain, except parietal lobe where higher amyloid burden in parietal lobe was associated with the presence of microbleed in the frontal region of the brain.

**Table 5:** Ordinary least square regression model demonstrating association between lobar microbleeds and amyloid-PET SUVr in global and lobar regions of the brain at baseline. Values given are effect size (95% confidence interval) p-value. Bold entries show significant associations.

| Amyloid PET SUVr | Microbleed present/absent? | | | | |
|---|---|---|---|---|---|
| | Whole cortex | Parietal lobe | Occipital lobe | Frontal lobe | Temporal lobe |
| Cortex | **0.04 (0.00 to 0.08) 0.041** | 0.05 (-0.01 to 0.11) 0.113 | 0.02 (-0.05 to 0.08) 0.620 | 0.05 (-0.02 to 0.12) 0.175 | 0.01 (-0.05 to 0.07) 0.825 |
| Temporal | **0.04 (0.00 to 0.07) 0.048** | 0.06 (-0.00 to 0.12) 0.053 | 0.01 (-0.05 to 0.08) 0.705 | 0.05 (-0.00 to 0.10) 0.068 | 0.00 (-0.06 to 0.06) 0.901 |
| Occipital | 0.03 (-0.00 to 0.06) 0.060 | 0.03 (-0.02 to 0.09) 0.196 | 0.04 (-0.02 to 0.10) 0.169 | 0.04 (-0.01 to 0.09) 0.109 | 0.01 (-0.04 to 0.06) 0.703 |
| Frontal | **0.04 (0.00 to 0.08) 0.047** | 0.05 (-0.02 to 0.12) 0.141 | 0.01 (-0.06 to 0.08) 0.754 | 0.02 (-0.03 to 0.06) 0.397 | 0.01 (-0.06 to 0.07) 0.809 |
| Parietal | **0.04 (0.00 to 0.08) 0.043** | 0.05 (-0.02 to 0.11) 0.155 | 0.02 (-0.05 to 0.09) 0.648 | **0.06 (0.00 to 0.12) 0.042** | 0.01 (-0.06 to 0.07) 0.803 |
| CENTILOIDS | **7.36 (1.09 to 13.63) 0.022** | 8.96 (-1.17 to 19.08) 0.083 | 4.11 (-6.87 to 15.09) 0.463 | 0.05 (-0.00 to 0.11) 0.070 | 3.17 (-7.04 to 13.37) 0.543 |

### 3.7 Lobar Microbleeds Are Associated with Amyloid Accumulation

Using ordinary least squares regression, after adjusting for age, gender, education, ApoE ε4 status, and clinical diagnosis, having at least one lobar microbleed was associated with increased odds of having abnormally higher CSF p-tau (coefficient = 2.85, 95% CI = 0.65 to 5.05, P = 0.011) and total tau levels (coefficient = 25.58, 95% CI = 5.31 to 45.85, P = 0.014). Similarly, after adjusting for age, gender, and education, having at least one lobar microbleed was associated with higher odds of having lower CSF Aβ levels (coefficient=-101.31, 95% CI=-167.73 to -34.88, P = 0.003).

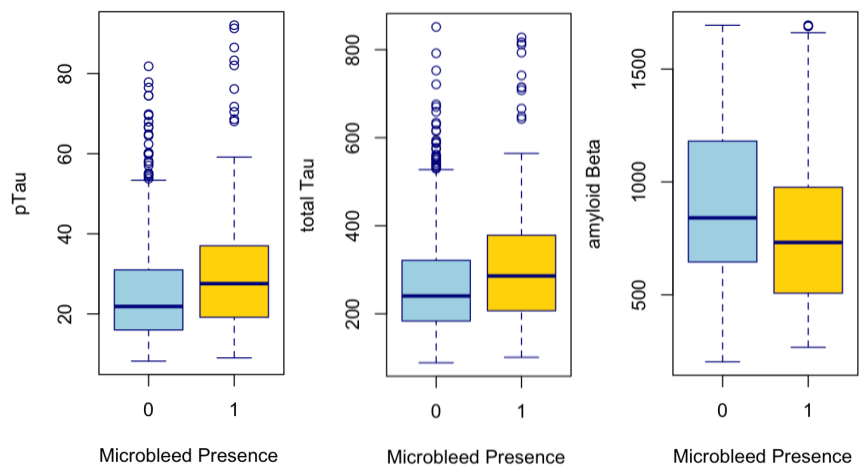

**Figure 2:** CSF proteomic markers in patients having microbleeds (1 on x-axis) and not having microbleeds (0 on x-axis).

## 4. Discussion

Our study evaluates associations between the presence of microbleeds in the overall brain cortex and at the lobar level with cognitive performance, and amyloid and tau pathology. Overall, the results of our analysis suggest a differential association of lobar microbleeds with Alzheimer pathogenesis, which reveals the importance of lobar microbleeds in prognostication.

The major finding from this analysis is that having at least one lobar microbleed is associated with accelerated longitudinal cognitive decline. Results discussed in the literature evaluating the association between microbleeds and cognitive decline have been mixed, mostly dependent on the studied cohort such as community-based populations, [24], [25], [26], [27] patients having a history of stroke or suspected stroke, [28], [29] or patients from a memory clinic who had MCI or AD, [30], [31], [32]. Also, cognition was being evaluated on a cross-sectional [24], [28], [29], [30], [31] or longitudinal, [25], [33], [34], [35], [36], [37] basis leading to variability in results across different studies. Most studies evaluating the association between microbleeds, and cognition demonstrated associations between microbleeds and either global or executive cognitive impairment at the cross-sectional level [28], [29], as well as increased decline in memory, greater longitudinal decline in Mini-Mental State Examination scores [32], higher probability of conversion from MCI to AD [25], and increased risk of developing dementia [37].

In this analysis, we aim to perform a comprehensive analysis of the effect that microbleeds may have on the clinical and pathological aspects of the disease. In our study, we found that the presence of microbleeds, either at the global or lobar level, was not associated with greater global cognitive impairment as measured by the ADAS-Cog11, and no significant associations were observed at the domain-specific level (Table 2). However, we did observe a significant association between lobar microbleeds and accelerated cognitive decline, which persisted after adjusting for age, gender, education, diagnosis, and APOE4 status (Table 3). This finding is consistent with a postmortem study by Arvanitakis et al. [38], which reported that moderate-to-severe amyloid angiopathy is linked to declines in perceptual speed and episodic memory, even after adjusting for concurrent AD pathology. When analyzed longitudinally, our results indicated that the presence of lobar microbleeds was associated with a greater decline in cognition. This finding aligns with previous studies that reported greater cognitive decline in patients having equal to or more than two [37] and five [26] lobar microbleeds. The association was particularly pronounced in the semantic and language domains, where cognitive decline in these areas depended on the location of microbleeds in different brain lobes. Although the associations were weaker in the language domain, they still showed a consistent, though modest, impact on cognitive decline. The differential effects of microbleed location on cognitive decline may suggest varying underlying etiologies of the microbleeds.

We evaluated the associations between the presence of lobar microbleeds and tau pathology at baseline. Our findings revealed that lobar microbleeds were significantly associated with increased tau PET signals in various brain regions. Notably, microbleeds in the temporal region were linked to a widespread increase in tau PET signals across different lobes of the brain. This phenomenon was not observed in other lobes, where the presence of microbleeds did not correlate with elevated tau PET signals. Microbleeds indicate underlying vascular pathology, such as cerebral amyloid angiopathy or other forms of small vessel disease. Vascular dysfunction can disrupt the blood-brain barrier, increasing its permeability and allowing harmful substances to enter brain tissue, promoting neuroinflammation. This inflammatory response can facilitate the spread and aggregation of tau protein beyond the initial site of vascular damage. We also found that the presence of microbleeds in specific lobar regions does not translate to pathology buildup in those regions rather lobes in temporal lobe effect tau-buildup in all the regions of the brain.

The analysis of associations with amyloid pathology demonstrated that the presence of lobar microbleeds is linked to increased and widespread brain amyloidosis, both globally and across various brain regions (Table 5). This finding aligns with earlier studies that reported reduced Aβ levels in CSF [39] and higher amyloid uptake on amyloid PET scans [40], [41] in patients having microbleeds compared to patients not having any microbleeds. These results further support the notion that lobar microbleeds are indicative of underlying amyloid angiopathy, as suggested by previous research [16], [18]. Amyloid angiopathy is present in 78%–98% of postmortem examinations of AD patients [42]. Moreover, the observation that only a higher burden of lobar microbleeds was associated with abnormal CSF Aβ levels implies that severe amyloid angiopathy may be more closely related to widespread brain amyloidosis and could play a role in advancing the disease. Moreover, when evaluating CSF, we found reduced Aβ levels in CSF and higher p-tau and total tau levels in patients with microbleed which is also consistent with previous literature [39], [43].

Our study has several important limitations. First, the study population was not racially or ethnically diverse, with over 90% of participants being white. Additionally, the cohort primarily consisted of individuals who were more educated and had fewer comorbidities than a typical community-based population of the same age [17]. This lack of diversity may limit the generalizability of the findings to more diverse or less healthy populations. Furthermore, in examining the associations between amyloid and tau pathology with microbleeds, we used the tau and amyloid PET scans closest to the microbleed assessment, which could sometimes be years apart. This time gap may introduce variability, as tau and amyloid pathology could have progressed significantly between the two time points. In future studies, we plan to use baseline scans collected within a defined timeframe and investigate how microbleeds affect longitudinal changes in tau and amyloid deposition across different brain regions. Additionally, participants with significant comorbid vascular disease, as indicated by a modified Hachinski score greater than 4, were excluded from ADNI. This exclusion limits our ability to detect associations that may be present in individuals with more severe vascular pathology. Future validation including individuals with greater vascular comorbidity and racial diversity, will be essential to confirm these associations.